\documentstyle[amssymb,preprint,aps]{revtex}

\begin{document}
\title{Quantum computation based on particle statistics}
\author{Giuseppe Castagnoli}
\address{Information Technology Dept., Elsag Bailey, 16154 Genova, Italy}
\author{Dalida Monti}
\address{Universit\`{a} di Genova and Elsag Bailey, 16154 Genova, Italy}
\date{\today }
\maketitle

\begin{abstract}
In spite of their evident logical character, particle statistics symmetries 
{\em are not} among the inherently quantum features exploited in quantum
computation. A difficulty may be that, being a constant of motion of a
unitary evolution, a particle statistics symmetry cannot affect the course
of such an evolution. We try to avoid this possible deadlock by introducing
a generalized (counterfactual, blunt) formulation where\ this type of
symmetry becomes a watchdog effects {\em shaping} the evolution of a unitary
computation process. The work is an exploration.
\end{abstract}

\section{Introduction}

This study deals with an alternative form of quantum computation. In this
context, it is useful to identify the ``boundary'' of the usual approach.

Until~now,~quantum~computation~speed~up$^{\text{[1-5 among others]}}$ (with
respect to the known classical algorithms) has taken advantage of a wide but
incomplete\ set\thinspace\ of inherently quantum features: effective
reversibility, superposition, entanglement, interference and measurement.
However, particle (fermionic, bosonic) statistics has never been exploited.

We should further note that all quantum algorithms found so far are based on
the same form of computation, which is {\em time-sequential}, in the sense
that a reversible Boolean network appears in the {\em time-diagram} of the
logical transformations undergone by the register.

They are also based on a common strategy, as recently recognized$^{\left[ 6-8%
\right] }$; for example, they can be seen as sophisticated applications of
the Mach-Zender interferometer.

Furtherly, this form of computation is suspected$^{\left[ 9\right] }$ to be
inherently unable of solving in polynomial time an NP-complete problem.%
\footnote{%
Solving in polynomial time (P) a problem formerly classified in NP (in
particular, whose solution required exponential time with all formerly known
algorithms), means declassifying to P\ that problem alone. Solving in
polynomial time an NP-complete problem would declassify to P\ all
NP-complete and all NP problems.}

The alternative approach explored in this study is tailored to solve a
particular NP-complete problem$^{\left[ 10\right] }$ and is essentially
based on the exploitation of particle statistics -- under a generalized,
counterfactual and {\em blunt} interpretation thereof.\ 

We give up time-sequential computation, by considering, in place of the
register, a reversible Boolean network fully deployed in space (rather than
appearing in the register time-diagram). Time is now orthogonal to the
network lay-out (fig. 1). In this way, part of the network input and part of
the output can both be constrained (fig. 1). Checking whether this network
is satisfiable is a version of the NP-complete satisfiability (SAT) problem.

We shall describe this in some detail; in the meantime, given the novelty of
the approach, we give an outline of the entire work.

\bigskip

\begin{center}
Fig. 1

\medskip

\begin{tabular}{c}
\hline
\multicolumn{1}{|c|}{\small b} \\ \hline
\multicolumn{1}{|c|}{\small 1} \\ \hline
{\small 4} \\ \hline
\multicolumn{1}{|c|}{\small f} \\ \hline
\multicolumn{1}{|c|}{\small 0} \\ \hline
{\small 5}
\end{tabular}
\quad 
\begin{tabular}{cccc}
\hline
\multicolumn{2}{|c}{\small I} & \multicolumn{2}{|c|}{\small O} \\ \hline
\multicolumn{1}{|c}{\small a} & \multicolumn{1}{|c}{\small b} & 
\multicolumn{1}{|c}{\small c} & \multicolumn{1}{|c|}{\small d} \\ \hline
\multicolumn{1}{|c}{\small 0} & \multicolumn{1}{|c}{\small 0} & 
\multicolumn{1}{|c}{\small 0} & \multicolumn{1}{|c|}{\small 0} \\ \hline
\multicolumn{1}{|c}{\small 0} & \multicolumn{1}{|c}{\small 1} & 
\multicolumn{1}{|c}{\small 0} & \multicolumn{1}{|c|}{\small 1} \\ \hline
\multicolumn{1}{|c}{\small 1} & \multicolumn{1}{|c}{\small 0} & 
\multicolumn{1}{|c}{\small 1} & \multicolumn{1}{|c|}{\small 1} \\ \hline
\multicolumn{1}{|c}{\small 1} & \multicolumn{1}{|c}{\small 1} & 
\multicolumn{1}{|c}{\small 1} & \multicolumn{1}{|c|}{\small 0} \\ \hline
\multicolumn{4}{c}{\small 1} \\ 
&  &  &  \\ 
&  &  & 
\end{tabular}
\quad 
\begin{tabular}{cc}
\hline
\multicolumn{1}{|c}{\small I} & \multicolumn{1}{|c|}{\small O} \\ \hline
\multicolumn{1}{|c}{\small d} & \multicolumn{1}{|c|}{\small e} \\ \hline
\multicolumn{1}{|c}{\small 0} & \multicolumn{1}{|c|}{\small 1} \\ \hline
\multicolumn{1}{|c}{\small 1} & \multicolumn{1}{|c|}{\small 0} \\ \hline
\multicolumn{2}{c}{\small 2} \\ 
&  \\ 
&  \\ 
&  \\ 
& 
\end{tabular}
\quad 
\begin{tabular}{cccc}
\hline
\multicolumn{2}{|c}{\small I} & \multicolumn{2}{|c|}{\small O} \\ \hline
\multicolumn{1}{|c}{\small e} & \multicolumn{1}{|c}{\small f} & 
\multicolumn{1}{|c}{\small g} & \multicolumn{1}{|c|}{\small h} \\ \hline
\multicolumn{1}{|c}{\small 0} & \multicolumn{1}{|c}{\small 0} & 
\multicolumn{1}{|c}{\small 0} & \multicolumn{1}{|c|}{\small 0} \\ \hline
\multicolumn{1}{|c}{\small 0} & \multicolumn{1}{|c}{\small 1} & 
\multicolumn{1}{|c}{\small 0} & \multicolumn{1}{|c|}{\small 1} \\ \hline
\multicolumn{1}{|c}{\small 1} & \multicolumn{1}{|c}{\small 0} & 
\multicolumn{1}{|c}{\small 1} & \multicolumn{1}{|c|}{\small 1} \\ \hline
\multicolumn{1}{|c}{\small 1} & \multicolumn{1}{|c}{\small 1} & 
\multicolumn{1}{|c}{\small 1} & \multicolumn{1}{|c|}{\small 0} \\ \hline
{\small 3} & \multicolumn{1}{c}{} & \multicolumn{1}{c}{} & 
\multicolumn{1}{c}{} \\ 
&  &  &  \\ 
&  &  & 
\end{tabular}
\quad 
\begin{tabular}{c}
\hline
\multicolumn{1}{|c|}{\small h} \\ \hline
\multicolumn{1}{|c|}{\small 1} \\ \hline
{\small 6}
\end{tabular}
\quad 
\begin{tabular}{cccccccc}
\hline
\multicolumn{1}{|c}{\small a} & \multicolumn{1}{|c}{\small b} & 
\multicolumn{1}{|c}{\small c} & \multicolumn{1}{|c}{\small d} & 
\multicolumn{1}{|c}{\small e} & \multicolumn{1}{|c}{\small f} & 
\multicolumn{1}{|c}{\small g} & \multicolumn{1}{|c|}{\small h} \\ \hline
\multicolumn{1}{|c}{\small 0} & \multicolumn{1}{|c}{\small 0} & 
\multicolumn{1}{|c}{\small 0} & \multicolumn{1}{|c}{\small 0} & 
\multicolumn{1}{|c}{\small 0} & \multicolumn{1}{|c}{\small 0} & 
\multicolumn{1}{|c}{\small 0} & \multicolumn{1}{|c|}{\small 0} \\ \hline
\multicolumn{1}{|c}{\small 0} & \multicolumn{1}{|c}{\small 0} & 
\multicolumn{1}{|c}{\small 0} & \multicolumn{1}{|c}{\small 0} & 
\multicolumn{1}{|c}{\small 0} & \multicolumn{1}{|c}{\small 0} & 
\multicolumn{1}{|c}{\small 0} & \multicolumn{1}{|c|}{\small 1} \\ \hline
\multicolumn{1}{|c}{\small .} & \multicolumn{1}{|c}{\small .} & 
\multicolumn{1}{|c}{\small .} & \multicolumn{1}{|c}{\small .} & 
\multicolumn{1}{|c}{\small .} & \multicolumn{1}{|c}{\small .} & 
\multicolumn{1}{|c}{\small .} & \multicolumn{1}{|c|}{\small .} \\ \hline
\multicolumn{1}{|c}{\small 1} & \multicolumn{1}{|c}{\small 1} & 
\multicolumn{1}{|c}{\small 1} & \multicolumn{1}{|c}{\small 1} & 
\multicolumn{1}{|c}{\small 1} & \multicolumn{1}{|c}{\small 1} & 
\multicolumn{1}{|c}{\small 1} & \multicolumn{1}{|c|}{\small 1} \\ \hline
\multicolumn{8}{c}{\small N} \\ 
\multicolumn{8}{c}{} \\ 
&  &  &  &  &  &  &  \\ 
&  &  &  &  &  &  & 
\end{tabular}
\end{center}

\begin{itemize}
\item  $a,b,...,h$ (fig.1)\ labeling the network nodes, are Boolean
variables. The overall network, or each gate, has its own input (left) and
output (right) nodes-variables: $a,b,f$\ ($c,g,h$) make the input (output)
of the overall network; $a,b$\ ($c,d$) make the input (output) of gate $\#1$%
, etc. With time orthogonal to the network lay-out, the terms ``input'' and
``output'' loose part of their meaning but are kept out of habit.

\item  Each gate introduces a logical constraint between its input and
output variables, by establishing a function between them. This function is
given in table form: the gate ``truth table''. For example table 1,
associated with gate 1, gives the invertible XOR (exclusive or) function.
Table 2, associated with gate 2, gives the NOT\ function (to spare notation,
the two-node NOT\ gate will be used as a network wire -- an inverting wire,
without loss of generality). Tables 4, 5, 6 represent the (partial) network
input/output constraints.

\item  Table $N$ is introduced for convenience; it does not establish any
constraint, its $2^{8}$\ rows are all the possible assignment of the network
Boolean variables.

\item  The SAT problem is whether there is at least one Boolean assignment
-- one row of $N$\ -- satisfying all the network constraints (all gates and
wires and the input/output constraints).

\item  The above is readily translated in Hilbert space mathematics$^{\left[
10\right] }$. A specific assignment of all network Boolean variables (one
row of $N$) becomes a tensor product of corresponding qubit eigenstates, and
a basis vector of the network Hilbert space ${\cal H}_{N}$. The logical
constraint established by each gate or wire (network element or N.E.) $i$
can be seen as a projector $A_{i}$ from ${\cal H}_{N}$ on ${\cal H}%
_{i}^{\left( c\right) }$. ${\cal H}_{i}^{\left( c\right) }$ is a
``constrained'' Hilbert subspace, spanned by those basis vectors of ${\cal H}%
_{N}$ which satisfy the constraint established by N.E. $i$.

\item  Qubits are implemented as a distributed system of two-state identical
particles. The particle distribution corresponding to the input/output nodes
of N.E. $i$ is subject to a Hamiltonian $H_{i}$. Given a suitable form of $%
H_{i}$, the projector $A_{i}$ is physically ``turned on'' in the degenerate
ground state of $H_{i}$, more generally in any network state satisfying $%
\left\langle \Psi \right| H_{i}\left| \Psi \right\rangle =0$. Such
projectors are shown to be epiphenomena of identical particle statistics,
under the generalized interpretation$^{\left[ 11,12\right] }$. All $H_{i}$
commute pairwise and the network Hamiltonian is $H_{N}=\sum_{i}H_{i}$.

\item  The network is prepared in a specific ground state $\left| \Psi
\left( 0\right) \right\rangle $ of $H_{N\text{ }}$\footnote{%
The generic ground state of $H_{N}$ is $2^{n}$ degenerate, where $n$ is the
number of input (or output) qubits.}. $\left\langle \Psi \left( 0\right)
\right| H_{N}\left| \Psi \left( 0\right) \right\rangle =0$ implies $%
\left\langle \Psi \left( 0\right) \right| H_{i}\left| \Psi \left( 0\right)
\right\rangle =0$ for all $i$, consequently all $A_{i}$ are ``turned on'' in 
$\left| \Psi \left( 0\right) \right\rangle $. Thus $\left| \Psi \left(
0\right) \right\rangle $ satisfies $\Pi _{i}A_{i}\left| \Psi \left( 0\right)
\right\rangle =\left| \Psi \left( 0\right) \right\rangle $, namely all N.E.
constraints (all $A_{i}$ are pairwise commuting). This preparation is also a
linear combination of all ${\cal H}_{N}$ basis vectors which satisfy the
network input constraints, not necessarily the output constraints. Since the
output constraints are removed in this phase, until now computation has
solved a simple P (polynomial) problem.

\item  By suitably operating on the output qubits in order to bring them in
match with their constraints, the network ground state, continuously
submitted to the projection $\Pi _{i}A_{i}\left| \Psi \left( t\right)
\right\rangle =\left| \Psi \left( t\right) \right\rangle $, is {\em unitarily%
} brought from $\left| \Psi \left( 0\right) \right\rangle $ to a state $%
\left| \Psi ^{^{\prime }}\left( \tau \right) \right\rangle $ satisfying also
the output constraints (if the network is satisfiable). This requires
interpreting a particle statistics symmetry as a continuous watchdog effect
shaping the network evolution \footnote{%
An example of a unitary evolution shaped by a watchdog effect, is the
evolution of the polarization of a photon going through an infinite series
of polarizing filters, each rotated by an infinitesimal constant angle with
respect to the former one. In a way, we go back to the root of quantum
computation (computation reversibility$^{\left[ \text{13,14}\right] }$) and
take an alternative branch, by exploring a strictly quantum form of
reversible computation.}. It should be noted that $\left| \Psi ^{^{\prime
}}\left( \tau \right) \right\rangle $ can be in superposition with an
orthogonal excited state $\left| \Psi ^{^{\prime \prime }}\left( \tau
\right) \right\rangle $, which is an error state from a computational
standpoint. However, the amplitude of this latter state should be small.

\item  Measuring the network after the above operations gives a Boolean
assignment. If the network is satisfiable, this is likely\ to be a network
solution (whether it is a solution is checkable off line in polynomial
time). If the check gives a non-solution, the network is likely to be not
satisfiable. By repeating the overall process for a sufficient number of
times, it can be ascertained with any confidence level whether the network
is satisfiable. Computation has now solved the hard, NP-complete, SAT
problem. By the way, the constrained network of fig. 1 is satisfied by
(only) $\left| \Psi \right\rangle =\left| 1\right\rangle _{a}\left|
1\right\rangle _{b}\left| 1\right\rangle _{c}\left| 0\right\rangle
_{d}\left| 1\right\rangle _{e}\left| 0\right\rangle _{f}\left|
1\right\rangle _{g}\left| 1\right\rangle _{h}$.

\item  The possibility that particle statistics yields a computation
speed-up appears to be promising in principle.
\end{itemize}

\section{A counterfactual interpretation of particle statistics}

This work hinges on a counterfactual interpretation of particle statistics$^{%
\left[ 11,12,15\right] }$. A particle statistics symmetry is usually
considered to be an initial condition which is conserved as a constant of
motion. This notion can be made more general by applying a counterfactual
logic: the state of the system {\em might} violate the symmetry, but{\em \
it cannot be so }because of a special watchdog effect which would
immediately project such a state on the symmetric subspace. This way of
reasoning yields to effective consequences\footnote{%
besides the fact that particle statistics symmetries should no more be put
among the initial conditions.}, diverging from the conventional application
of quantum mechanics. Apropos of this, it is worth citing the following
statement by Roger Penrose: ``What is particularly curious about quantum
theory is that there can be actual physical effects arising from what
philosophers refer to as {\em counterfactuals} $-$ that is things that might
have happened, although did not in fact happen''$^{\left[ \text{16}\right]
}. $\footnote{%
The relevance of counterfactual reasoning in {\em sequential} computation
has been highlighted by Richard Jozsa$^{\left[ 17\right] }$.}

By way of exemplification, let us consider a state which we assume to be
symmetrical (for particle permutation) because of particle statistics:

\begin{equation}
\left| \Psi \left( 0\right) \right\rangle =\cos ^{2}\vartheta \left|
0\right\rangle _{1}\left| 0\right\rangle _{2}+\sin \vartheta \cos \vartheta
\left( \left| 0\right\rangle _{1}\left| 1\right\rangle _{2}+\left|
1\right\rangle _{1}\left| 0\right\rangle _{2}\right) +\sin ^{2}\vartheta
\left| 1\right\rangle _{1}\left| 1\right\rangle _{2};
\end{equation}
1 and 2 label two identical two-state particles (which makes this an
idealized situation). The above amplitudes are used for convenience with no
significant restriction. Of course $\left| \Psi \left( 0\right)
\right\rangle $ is symmetrical under the particle permutation $P_{12}$ iff
it satisfies the equation $S_{12}\left| \Psi \left( 0\right) \right\rangle
=\left| \Psi \left( 0\right) \right\rangle $, where $S_{12}=\frac{1}{2}%
\left( 1+P_{12}\right) $ is the symmetrization projector. Let ${\cal H}_{12}$
be the Hilbert space of the two particles and ${\cal H}_{12}^{\left(
s\right) }$ be the symmetric subspace:

\[
{\cal H}_{12}^{\left( s\right) }=span\left\{ \left| 0\right\rangle
_{1}\left| 0\right\rangle _{2},\frac{1}{\sqrt{2}}\left( \left|
0\right\rangle _{1}\left| 1\right\rangle _{2}+\left| 1\right\rangle
_{1}\left| 0\right\rangle _{2}\right) ,\left| 1\right\rangle _{1}\left|
1\right\rangle _{2}\right\} . 
\]

In order to introduce the counterfactual picture, we assume that $\left|
\Psi \left( t\right) \right\rangle $ {\em might} violate the symmetry. This
is interpreted by taking, for all $t$, a $\left| \Psi \left( t\right)
\right\rangle $ which is {\em a free normalized vector }of ${\cal H}_{12}$.
Namely, given:

\begin{equation}
\left| \Psi \left( t\right) \right\rangle =\sum_{i,j\in \left\{ 0,1\right\}
}\alpha _{ij}^{\left( t\right) }\left| i\right\rangle _{1}\left|
j\right\rangle _{2},\text{ with }\sum_{i,j\in \left\{ 0,1\right\} }\left|
\alpha _{ij}^{\left( t\right) }\right| ^{2}=1,
\end{equation}

\noindent the $\alpha _{ij}^{\left( t\right) }$\ are free complex variables
independent of each other up to normalization for any time $t$. In other
words, if $t_{1}\neq t_{2}$, $\left| \Psi \left( t_{1}\right) \right\rangle $
and $\left| \Psi \left( t_{2}\right) \right\rangle $ are two {\em independent%
} free normalized vectors of ${\cal H}_{12}$.

The logic is that, at any time $t=t_{-}$, $\left| \Psi \left( t_{-}\right)
\right\rangle $ might violate the symmetry: it is thus {\em a free vector }%
of ${\cal H}_{12}$. Then $\left| \Psi \left( t_{-}\right) \right\rangle $
should be projected on ${\cal H}_{12}^{\left( s\right) }$. This gives as a
result the vector of ${\cal H}_{12}^{\left( s\right) }$ closest to $\left|
\Psi \left( t_{-}\right) \right\rangle .$

Let $\left| \Psi \left( t_{+}\right) \right\rangle $ be a free normalized
vector of ${\cal H}_{12}$ independent from $\left| \Psi \left( t_{-}\right)
\right\rangle $ $\left( t_{+}>t_{-}\right) $. The above projection can be
obtained by submitting $\left| \Psi \left( t_{+}\right) \right\rangle $ to
the following conditions:

\medskip for all $t$:

\begin{enumerate}
\item[i)]  $S_{12}\left| \Psi \left( t_{+}\right) \right\rangle =\left| \Psi
\left( t_{+}\right) \right\rangle ,$

\item[ii)]  $\left| \left\langle \Psi \left( t_{+}\right) \right. \left| \
\Psi \left( t_{-}\right) \right\rangle \right| $ must be maximum.
\end{enumerate}

\noindent In this picture, $\left| \Psi \left( t_{-}\right) \right\rangle $
is the vector ``immediately before projection'' while $\left| \Psi \left(
t_{+}\right) \right\rangle $, subject to conditions (i) and (ii), is the
vector ``immediately after projection''. This\ counterfactual logic will
become clearer when applied to an evolution of the above triplet state.

Continuous projection can be interpreted as a {\em continuous} {\em state
vector reduction} induced by particle statistics. To show this, it is
convenient to rewrite (2) as follows:

\begin{eqnarray*}
\left| \Psi \left( t_{-}\right) \right\rangle &=&\alpha _{00}^{\left(
t_{-}\right) }\left| 0\right\rangle _{1}\left| 0\right\rangle _{2}+\frac{1}{2%
}\left( \alpha _{01}^{\left( t_{-}\right) }+\alpha _{10}^{\left(
t_{-}\right) }\right) \left( \left| 0\right\rangle _{1}\left| 1\right\rangle
_{2}+\left| 1\right\rangle _{1}\left| 0\right\rangle _{2}\right) + \\
&&\frac{1}{2}\left( \alpha _{01}^{\left( t_{-}\right) }-\alpha _{10}^{\left(
t_{-}\right) }\right) \left( \left| 0\right\rangle _{1}\left| 1\right\rangle
_{2}-\left| 1\right\rangle _{1}\left| 0\right\rangle _{2}\right) +\alpha
_{11}^{\left( t_{-}\right) }\left| 1\right\rangle _{1}\left| 1\right\rangle
_{2}.
\end{eqnarray*}
This vector undergoes a {\em partial }\footnote{%
It is essential that ${\cal H}_{12}^{\left( s\right) }$ has dimension higher
than one.} state vector reduction on the subspace ${\cal H}_{12}^{\left(
s\right) }$. This cancels the amplitude of the channel ($\left|
0\right\rangle _{1}\left| 1\right\rangle _{2}-\left| 1\right\rangle
_{1}\left| 0\right\rangle _{2}$) and renormalizes the amplitudes of the
other channels. This can be seen as a {\em special form of} {\em %
interference }(from now on: ``interference''), destructive in amplitude
cancelation and constructive in renormalization. Of course this form of
state vector reduction is an interpretation of particle statistics and is
not due to an interaction with an external apparatus.

$t_{+}$\ might be as close to\ $t_{-}$\ as desired, since state vector
reduction does not involve the notion of elapsed time. However, we will take 
$t_{-}=t-dt$\ and $t_{+}=t$, in view of applying (i) and (ii) to an
evolution described by $\left| \Psi \left( t\right) \right\rangle $. The
former conditions are thus rewritten:\medskip

for all $t$:

\begin{enumerate}
\item[i)]  $S_{12}\left| \Psi \left( t\right) \right\rangle =\left| \Psi
\left( t\right) \right\rangle ,$

\item[ii)]  $Max$ $\left| \left\langle \Psi \left( t\right) \right. \left| \
\Psi \left( t-dt\right) \right\rangle \right| ,$

where $\left| \Psi \left( t\right) \right\rangle $ is a free normalized
vector of ${\cal H}_{12}$.
\end{enumerate}

Now we go back to our system in a triplet state. Duly constrained, $\left|
\Psi \left( t\right) \right\rangle $ will describe an evolution of that
state. Conditions (i) and (ii) will be applied {\em together} with an
operation performed on the system, say an identical rotation of the state of
both particles of an angle $\varphi \left( t\right) $ (where $\varphi \left(
t\right) $ is some function of time), starting from preparation (1).

Let us anticipate the result, by following the conventional approach. The
above operation, symmetrical for particles permutation, commutes with $%
S_{12} $, therefore symmetry is a constant of motion of the evolution;
namely,\ it does not apply any constraint to it. The two particles can thus
be rotated independently of each other. This yields of course:

\begin{eqnarray}
\left| \Psi \left( t\right) \right\rangle &=&\cos ^{2}\left[ \vartheta
+\varphi \left( t\right) \right] \left| 0\right\rangle _{1}\left|
0\right\rangle _{2}+\sin \left[ \vartheta +\varphi \left( t\right) \right]
\cos \left[ \vartheta +\varphi \left( t\right) \right] \left( \left|
0\right\rangle _{1}\left| 1\right\rangle _{2}+\left| 1\right\rangle
_{1}\left| 0\right\rangle _{2}\right) \\
&&+\sin ^{2}\left[ \vartheta +\varphi \left( t\right) \right] \left|
1\right\rangle _{1}\left| 1\right\rangle _{2}  \nonumber
\end{eqnarray}

Now we want to obtain the same result by {\em ignoring} the fact that
symmetry is conserved as a constant of motion, and by resorting to the
projection interpretation of particle statistics, that is to conditions (i)
and (ii). We need to add a further condition to specify that the state of
each particle has been rotated by $\varphi \left( t\right) $ with respect to
the initial state (1).

Doing this requires some attentions. Now the two particles cannot be
considered independently of each other, since their overall state is
submitted to the continuous projection, or watchdog effect, (i) and (ii). We
must utilize the particle density matrices:

\[
\rho _{i}\left( t\right) =Tr_{3-i}\left[ \left| \Psi \left( t\right)
\right\rangle \left\langle \Psi \left( t\right) \right| \right] , 
\]

\noindent where $Tr_{3-i}$ means partial trace over $3-i$, and $i=1,2$ is
the particle label (if we use the method of random phases$^{\left[ 18\right]
}$, $\left| \Psi \left( t\right) \right\rangle $ does not need to be a pure
state -- anyhow it will turn out to be that). Furthermore, the coherence
elements of each density matrix -- as overall entanglement -- can be
affected by the watchdog effect (in fact they will be determined by it). We
only know that the diagonal of each density matrix must show a $\varphi
\left( t\right) $ rotation. In conclusion the further condition (iii) is:

\[
diag\rho _{1}\left( t\right) =diag\left\{ Tr_{2}\left[ \left| \Psi \left(
t\right) \right\rangle \left\langle \Psi \left( t\right) \right| \right]
\right\} =\cos ^{2}\left[ \vartheta +\varphi \left( t\right) \right] \left|
0\right\rangle _{2}\left\langle 0\right| _{2}+\sin ^{2}\left[ \vartheta
+\varphi \left( t\right) \right] \left| 1\right\rangle _{2}\left\langle
1\right| _{2}. 
\]

\[
diag\rho _{2}\left( t\right) =diag\left\{ Tr_{1}\left[ \left| \Psi \left(
t\right) \right\rangle \left\langle \Psi \left( t\right) \right| \right]
\right\} =\cos ^{2}\left[ \vartheta +\varphi \left( t\right) \right] \left|
0\right\rangle _{2}\left\langle 0\right| _{2}+\sin ^{2}\left[ \vartheta
+\varphi \left( t\right) \right] \left| 1\right\rangle _{2}\left\langle
1\right| _{2}. 
\]

It is readily seen that the {\em simultaneous} application of conditions (i)
and (iii) yields:

\begin{eqnarray*}
\left| \Psi \left( t\right) \right\rangle &=&\cos ^{2}\left[ \vartheta
+\varphi \left( t\right) \right] \left| 0\right\rangle _{1}\left|
0\right\rangle _{2}+e^{i\delta _{1}}\sin \left[ \vartheta +\varphi \left(
t\right) \right] \cos \left[ \vartheta +\varphi \left( t\right) \right]
\left( \left| 0\right\rangle _{1}\left| 1\right\rangle _{2}+\left|
1\right\rangle _{1}\left| 0\right\rangle _{2}\right) \\
&&+e^{i\delta _{2}}\sin ^{2}\left[ \vartheta +\varphi \left( t\right) \right]
\left| 1\right\rangle _{1}\left| 1\right\rangle _{2},
\end{eqnarray*}

\noindent where the phases $\delta _{1}$ and $\delta _{2}$ are still
unconstrained; condition (ii), given the initial state (1), sets $\delta
_{1}=\delta _{2}=0$ and keeps this result frozen throughout the evolution,
as readily checked. This yields the desired evolution (3).

As readily seen, {\em by removing either one of the two conditions }(iii),
evolution (3) is still obtained -- the two rotations are redundant with
respect to one another.

Therefore it is perfectly legitimate to say that the {\em rotation of the
state of only} {\em one particle}, either one in an indistinguishable way, 
{\em drags an identical rotation of the state of the other}.

It should be noted that applying condition (iii) to either part of the
system does not mean applying the identity to the other part\footnote{%
Which would mean: (a) applying an infinitesimal rotation to one particle and
the identity to the other, (b) symmetrizing the result and (c) repeating the
cycle. This would bring in a Zeno effect freezing the evolution in its
initial state (1) -- see ref. [12].}. In the current formulation, the
transformation of this latter part must be {\em disregarded}, since it is
already determined by the transformation of the former part {\em and} the
watchdog effect. In a sense, either part, being indistinguishable,
represents the whole. This might seem a peculiar way of applying quantum
mechanics but it is tautologically true here.

In this idealized picture (the two-state particles), particle statistics
symmetry can be seen as an interaction-free constraint (working by way of
``interference'') between the states of the two particles, {\em before}
being a constant of motion. Of course symmetry is conserved by the
propagation (3), but this is due to the fact that the propagation is shaped
in a symmetric way by the watchdog effect described by (i) and (ii).

This work explores the possibility of porting the above, generalized,
interpretation of particle statistics to the case of {\em identical}
particles hosted by {\em distinguishable} sites, provided that the whole
system is in the ground state of a suitable Hamiltonian. Section III gives a
purely mathematical porting. The physical model will be the subject of
Section IV.

\section{The NOT gate as a projector}

We consider a simple NOT gate (fig. 1, gate and table \#2). When used as a
network (inverting) wire it will be called ``link''.

Mathematically, the previous model is readily ported to a pair of qubits $r$
and $s$ [$rs$ would run over $de$, etc. in a generic network], of Hilbert
space 
\[
{\cal H}_{rs}=span\{\left| 0\right\rangle _{r}\left| 0\right\rangle
_{s},\left| 0\right\rangle _{r}\left| 1\right\rangle _{s},\left|
1\right\rangle _{r}\left| 0\right\rangle _{s},\left| 1\right\rangle
_{r}\left| 1\right\rangle _{s}\}. 
\]

These two qubits will be implemented in a way that makes them
distinguishable (Section IV).

The constrained subspace, spanned by the basis vectors of ${\cal H}_{rs}$
which satisfy the NOT\ gate, is: 
\[
{\cal H}_{rs}^{\left( c\right) }=span\left\{ \left| 0\right\rangle
_{r}\left| 1\right\rangle _{s},\left| 1\right\rangle _{r}\left|
0\right\rangle _{s}\right\} . 
\]

The projector $A_{rs}$ from ${\cal H}_{rs}$ on ${\cal H}_{rs}^{\left(
c\right) }$ is defined by the following equations:

\begin{eqnarray*}
A_{rs}\left| 0\right\rangle _{r}\left| 1\right\rangle _{s} &=&\left|
0\right\rangle _{r}\left| 1\right\rangle _{s},\text{ \qquad }A_{rs}\left|
1\right\rangle _{r}\left| 0\right\rangle _{s}=\left| 1\right\rangle
_{r}\left| 0\right\rangle _{s}, \\
A_{rs}\left| 0\right\rangle _{r}\left| 0\right\rangle _{s} &=&0,\qquad \text{
}A_{rs}\left| 1\right\rangle _{r}\left| 1\right\rangle _{s}=0.
\end{eqnarray*}
Finding a physical model of $A_{rs}$ will be the subject of the next section.

The two qubits should be prepared in an initial state already belonging to $%
{\cal H}_{rs}^{\left( c\right) }$:

\begin{equation}
\left| \Psi \left( 0\right) \right\rangle =\cos \vartheta \left|
0\right\rangle _{r}\left| 1\right\rangle _{s}+\sin \vartheta \left|
1\right\rangle _{r}\left| 0\right\rangle _{s}.
\end{equation}

\noindent The evolution we are looking for, supposedly obtained by operating
only on qubit $r$ and disregarding qubit $s$ (see Section IV), is the
solution of the following system of simultaneous equations:

For all $t$:

\begin{enumerate}
\item[i)]  \smallskip $A_{rs}\left| \Psi \left( t\right) \right\rangle
=\left| \Psi \left( t\right) \right\rangle ,$

\item[ii)]  Max $\left| \left\langle \Psi \left( t\right) \right. \left|
\Psi \left( t-dt\right) \right\rangle \right| ,$
\end{enumerate}

\[
\text{iii) }diag\rho _{r}\left( t\right) =diag\left\{ Tr_{s}\left[ \left|
\Psi \left( t\right) \right\rangle \left\langle \Psi \left( t\right) \right| %
\right] \right\} =\cos ^{2}\left[ \vartheta +\varphi \left( t\right) \right]
\left| 0\right\rangle _{r}\left\langle 0\right| _{r}+\sin ^{2}\left[
\vartheta +\varphi \left( t\right) \right] \left| 1\right\rangle
_{r}\left\langle 1\right| _{r}, 
\]
\noindent where $\left| \Psi \left( t\right) \right\rangle $ is a free
normalized vector of ${\cal H}_{rs}:\left| \Psi \left( t\right)
\right\rangle =\sum_{i,j\in \left\{ 0,1\right\} }\alpha _{ij}^{\left(
t\right) }\left| i\right\rangle _{r}\left| j\right\rangle _{s},$ with $%
\sum_{i,j\in \left\{ 0,1\right\} }\left| \alpha _{ij}^{\left( t\right)
}\right| ^{2}=1$.

As readily seen, conditions (i) and (iii) yield:

\[
\left| \Psi \left( t\right) \right\rangle =\cos \left[ \vartheta +\varphi
\left( t\right) \right] \left| 0\right\rangle _{r}\left| 1\right\rangle
_{s}+e^{i\delta }\sin \left[ \vartheta +\varphi \left( t\right) \right]
\left| 1\right\rangle _{r}\left| 0\right\rangle _{s}, 
\]
where $\delta $ is an unconstrained phase; condition (ii), given the link
initial state (4), sets $\delta =0$ and keeps it frozen throughout the
evolution. This yields to the unitary evolution:

\begin{equation}
\left| \Psi \left( t\right) \right\rangle =\cos \left[ \vartheta +\varphi
\left( t\right) \right] \left| 0\right\rangle _{r}\left| 1\right\rangle
_{s}+\sin \left[ \vartheta +\varphi \left( t\right) \right] \left|
1\right\rangle _{r}\left| 0\right\rangle _{s}.
\end{equation}
We must assume that condition (iii) represents some physical operation
performed {\em only} on qubit $r$, under conditions (i) and (ii) -- exactly
like in Section II. The plausibility of this assumption will be discussed in
Section IV. This would mean that a transformation operated on qubit $r$
drags an identical transformation of qubit $s$. Mathematically, this is true:

\[
\rho _{s}\left( t\right) =T_{r_{r}}\left[ \left| \Psi \left( t\right)
\right\rangle \left\langle \Psi \left( t\right) \right| \right] =\sin ^{2}%
\left[ \vartheta +\varphi \left( t\right) \right] \left| 0\right\rangle
_{s}\left\langle 0\right| _{s}+\cos ^{2}\left[ \vartheta +\varphi \left(
t\right) \right] \left| 1\right\rangle _{s}\left\langle 1\right| _{s}; 
\]

\noindent with respect to condition (iii), 0 and 1 are interchanged because
qubit $s$ is the NOT of qubit $r$.

Interestingly, if all ${\cal H}_{rs}^{\left( c\right) }$ basis vectors occur
with amplitudes different from zero in the initial state (4), namely if $%
\vartheta \neq 0,\frac{\pi }{2}$, condition (i) is redundant with respect to
condition (ii). In this case, condition (ii) alone implies $\alpha
_{00}^{\left( t\right) }=\alpha _{11}^{\left( t\right) }=0$, and this
already satisfies condition (i).

On the contrary, condition (i) is not redundant if $\vartheta =0$ or $\frac{%
\pi }{2}$. For example, if $\vartheta =0$, i.e. $\left| \Psi \left( 0\right)
\right\rangle =\left| 0\right\rangle _{r}\left| 1\right\rangle _{s}$,
condition (ii) implies $\alpha _{00}^{\left( t\right) }=0$ and $\left|
\alpha _{01}^{\left( t\right) }\right| ^{2}=\cos ^{2}\left[ \vartheta
+\varphi \left( t\right) \right] $, as needed, while $\alpha _{10}^{\left(
t\right) }$ and $\alpha _{11}^{\left( t\right) }$ are only subject to the
constraint $\left| \alpha _{10}^{\left( t\right) }\right| ^{2}+\left| \alpha
_{11}^{\left( t\right) }\right| ^{2}=\sin ^{2}\left[ \vartheta +\varphi
\left( t\right) \right] $. Thus, disregarding condition (i) would allow for
the existence of the forbidden state $\alpha _{11}^{\left( t\right) }\left|
1\right\rangle _{r}\left| 1\right\rangle _{s}$.

It is worth noting that the same evolution (5) can be obtained by applying
the unitary operator $Q_{rs}\left[ \varphi \left( t\right) \right] $ to the 
{\em overall state} $\left| \Psi \left( t\right) \right\rangle $\footnote{%
the generator of $Q_{rs}\left[ \varphi \left( t\right) \right] $ commutes
with $H_{rs}$, the interaction Hamiltonian between qubits $r$ and $s$ which
will be introduced in the next Section.}:

$Q_{rs}\left[ \varphi \left( t\right) \right] \equiv \left( 
\begin{array}{cccc}
1 & 0 & 0 & 0 \\ 
0 & \cos \varphi \left( t\right) & \sin \varphi \left( t\right) & 0 \\ 
0 & -\sin \varphi \left( t\right) & \cos \varphi \left( t\right) & 0 \\ 
0 & 0 & 0 & 1
\end{array}
\right) $,

with $\left| 0\right\rangle _{r}\left| 1\right\rangle _{s}\equiv \left( 
\begin{array}{l}
1 \\ 
0
\end{array}
\right) _{r}\otimes \left( 
\begin{array}{l}
0 \\ 
1
\end{array}
\right) _{s},$ $\left| 1\right\rangle _{r}\left| 0\right\rangle _{s}\equiv
\left( 
\begin{array}{l}
0 \\ 
1
\end{array}
\right) _{r}\otimes \left( 
\begin{array}{l}
1 \\ 
0
\end{array}
\right) _{s}.$

\noindent $Q_{rs}\left[ \varphi \left( t\right) \right] $ operates on the
overall state in a non factorizable way, since at any time $t$ the
transformation undergone by qubit $r$ ($s$) is {\em conditioned} by the
state of qubit $s$ ($r$), because of entanglement:

\[
Q_{rs}\left[ \varphi \left( t\right) \right] \left| 0\right\rangle
_{r}\left| 1\right\rangle _{s}=\cos \varphi \left( t\right) \left|
0\right\rangle _{r}\left| 1\right\rangle _{s}+\sin \varphi \left( t\right)
\left| 1\right\rangle _{r}\left| 0\right\rangle _{s}, 
\]

\[
Q_{rs}\left[ \varphi \left( t\right) \right] \left| 1\right\rangle
_{r}\left| 0\right\rangle _{s}=-\sin \varphi \left( t\right) \left|
0\right\rangle _{r}\left| 1\right\rangle _{s}+\cos \varphi \left( t\right)
\left| 1\right\rangle _{r}\left| 0\right\rangle _{s}. 
\]

\noindent There is so to speak a ``hidden'' conditional logic: finding this
evolution $-$ i.e. $Q_{rs}\left[ \varphi \left( t\right) \right] $ $-$
amounts to solving a logical problem. This becomes NP-complete when the link
or the NOT\ gate belongs to a generic Boolean network.

We have thus ascertained a peculiar fact. Our ``operation on a part'' [this
is just the mathematical condition (iii) for the time being], {\em blind} to
its effect on the whole, performed together with continuous $A_{rs}$
projection, generates a {\em unitary} {\em transformation} which is, so to
speak, {\em wise} to the whole state, to how it should be transformed
without ever violating $A_{rs}$ (i.e. the link or the NOT\ gate). Of course $%
A_{rs}$ ends up commuting with the resulting overall unitary propagator, but
because this is {\em shaped} by $A_{rs}$.

\section{Turning on the NOT\ gate}

$A_{rs}$ projection can be shown to be an epiphenomenon of particle
(fermionic or bosonic) statistics ``turned on'' in a special physical
situation. In the following, we will adopt fermionic statistics.

We submit a couple of identical fermions 1 and 2 to a suitable Hamiltonian.
Each fermion must have two compatible, binary degrees of freedom $\chi $ and 
$\lambda $. Just for the sake of visualization (things should remain more
abstract), we can think that each fermion is a spin $1/2$ particle which can
occupy one of either two sites of a spatial lattice. $\chi $ can be the
particle spin (say that $\chi =0,1$ correspond to spin $=${\em down, up)}
and $\lambda =r,s$ the label of the site occupied by the particle. For
example, $\left| 0\right\rangle _{1}\left| 1\right\rangle _{2}\left|
r\right\rangle _{1}\left| s\right\rangle _{2}$ reads: particle $1$ spin $=$
down (0), particle $2$ spin $=$ up (1), particle $1$ site $=r,$ particle $2$
site $=s$. There are 16 combinations like this, which make up the basis of
the Hilbert space ${\cal H}_{12}$. However, there are only six {\em %
antisymmetrical} combinations (not violating fermion statistics) which make
up the basis of the {\em antisymmetrical} subspace ${\cal H}_{12}^{\left(
a\right) }$.

These basis vectors are represented in first and second quantization and,
when there is exactly one particle per site, in qubit notation ($\chi $ and $%
\lambda $ stand respectively for the qubit eigenvalue and label), $\left|
0\right\rangle $ is the vacuum vector:

$\left| a\right\rangle =\frac{1}{\sqrt{2}}\left( \left| 0\right\rangle
_{1}\left| 1\right\rangle _{2}-\left| 1\right\rangle _{1}\left|
0\right\rangle _{2}\right) \left| r\right\rangle _{1}\left| r\right\rangle
_{2}=a_{0r}^{\dagger }\ a_{1r}^{\dagger }\left| 0\right\rangle ,$

$\left| b\right\rangle =\frac{1}{\sqrt{2}}\left( \left| 0\right\rangle
_{1}\left| 1\right\rangle _{2}-\left| 1\right\rangle _{1}\left|
0\right\rangle _{2}\right) \left| s\right\rangle _{1}\left| s\right\rangle
_{2}=a_{0s}^{\dagger }\ a_{1s}^{\dagger }\left| 0\right\rangle ;$

$\left| c\right\rangle =\frac{1}{\sqrt{2}}\left| 0\right\rangle _{1}\left|
0\right\rangle _{2}\left( \left| r\right\rangle _{1}\left| s\right\rangle
_{2}-\left| s\right\rangle _{1}\left| r\right\rangle _{2}\right)
=a_{0r}^{\dagger }\ a_{0s}^{\dagger }\left| 0\right\rangle =\left|
0\right\rangle _{r}\left| 0\right\rangle _{s},$

$\left| d\right\rangle =\frac{1}{\sqrt{2}}\left| 1\right\rangle _{1}\left|
1\right\rangle _{2}\left( \left| r\right\rangle _{1}\left| s\right\rangle
_{2}-\left| s\right\rangle _{1}\left| r\right\rangle _{2}\right)
=a_{1r}^{\dagger }\ a_{1s}^{\dagger }\left| 0\right\rangle =\left|
1\right\rangle _{r}\left| 1\right\rangle _{s},$

$\left| e\right\rangle =\frac{1}{2}\left( \left| 0\right\rangle _{1}\left|
1\right\rangle _{2}+\left| 1\right\rangle _{1}\left| 0\right\rangle
_{2}\right) \left( \left| r\right\rangle _{1}\left| s\right\rangle
_{2}-\left| s\right\rangle _{1}\left| r\right\rangle _{2}\right) =$

$\frac{1}{\sqrt{2}}\left( a_{0r}^{\dagger }\ a_{1s}^{\dagger
}+a_{1r}^{\dagger }\ a_{0s}^{\dagger }\right) \left| 0\right\rangle =\frac{1%
}{\sqrt{2}}\left( \left| 0\right\rangle _{r}\left| 1\right\rangle
_{s}+\left| 1\right\rangle _{r}\left| 0\right\rangle _{s}\right) .$

$\left| f\right\rangle =\frac{1}{2}\left( \left| 0\right\rangle _{1}\left|
1\right\rangle _{2}-\left| 1\right\rangle _{1}\left| 0\right\rangle
_{2}\right) \left( \left| r\right\rangle _{1}\left| s\right\rangle
_{2}+\left| s\right\rangle _{1}\left| r\right\rangle _{2}\right) =$

$\frac{1}{\sqrt{2}}\left( a_{0r}^{\dagger }\ a_{1s}^{\dagger
}-a_{1r}^{\dagger }\ a_{0s}^{\dagger }\right) \left| 0\right\rangle =\frac{1%
}{\sqrt{2}}\left( \left| 0\right\rangle _{r}\left| 1\right\rangle
_{s}-\left| 1\right\rangle _{r}\left| 0\right\rangle _{s}\right) .$

\noindent $a_{\chi \lambda }^{\dagger }$ creates a particle of spin $\chi $
in site $\lambda $; creation/annihilation operators are subject to: $\left\{
a_{i}^{\dagger },a_{j}^{\dagger }\right\} =\left\{ a_{i},a_{j}\right\} =0,\
\ \left\{ a_{i}^{\dagger },a_{j}\right\} =\delta _{i,j}.$ Under the
condition that there is exactly one particle per site, they generate a qubit
algebra.

Now we introduce the Hamiltonian

\[
H_{rs}=E_{a}\left| a\right\rangle \left\langle a\right| +E_{b}\left|
b\right\rangle \left\langle b\right| +E_{c}\left| c\right\rangle
\left\langle c\right| +E_{d}\left| d\right\rangle \left\langle d\right| 
\]

\noindent or,~in~second~quantization, 
\[
H_{rs}=-(E_{a}\ a_{0r}^{\dagger }\ a_{1r}^{\dagger }a_{0r}a_{1r}+E_{b}\
a_{0s}^{\dagger }\ a_{1s}^{\dagger }a_{0s}a_{1s}+E_{c}\ a_{0r}^{\dagger }\
a_{0s}^{\dagger }a_{0r}a_{0s}+E_{d}\ a_{1r}^{\dagger }\ a_{1s}^{\dagger
}a_{1r}a_{1s}), 
\]
with $E_{a}$, $E_{b}$, $E_{c}$, $E_{d}\geq E$ discretely above 0. This
leaves us with two degenerate ground eigenstates:

\[
\left| e\right\rangle =\frac{1}{\sqrt{2}}\left( \left| 0\right\rangle
_{r}\left| 1\right\rangle _{s}+\left| 1\right\rangle _{r}\left|
0\right\rangle _{s}\right) \text{ and }\left| f\right\rangle =\frac{1}{\sqrt{%
2}}\left( \left| 0\right\rangle _{r}\left| 1\right\rangle _{s}-\left|
1\right\rangle _{r}\left| 0\right\rangle _{s}\right) . 
\]
The generic ground state is thus:

\begin{equation}
\left| \Psi \right\rangle =\alpha \left| 0\right\rangle _{r}\left|
1\right\rangle _{s}+\beta \left| 1\right\rangle _{r}\left| 0\right\rangle
_{s}\text{, with }\left| \alpha \right| ^{2}+\left| \beta \right| ^{2}=1.
\end{equation}
Of course $\left| \Psi \right\rangle $ satisfies $A_{rs}\left| \Psi
\right\rangle =\left| \Psi \right\rangle $, and belongs to ${\cal H}%
_{rs}^{\left( c\right) }$ (Section III), a subspace of ${\cal H}%
_{12}^{\left( a\right) }$.

Let $A_{12}\left| \Psi \right\rangle =\frac{1}{2}$ $\left( 1-P_{12}\right) $
be the usual antisymmetrization projector. Due to the above
anticommutation~relations: 
\[
A_{12}\left| 0\right\rangle _{r}\left| 1\right\rangle _{s}=\left|
0\right\rangle _{r}\left| 1\right\rangle _{s}\text{ and }A_{12}\left|
1\right\rangle _{r}\left| 0\right\rangle _{s}=\left| 1\right\rangle
_{r}\left| 0\right\rangle _{s},\text{ moreover} 
\]
\[
A_{12}\left| 0\right\rangle _{r}\left| 0\right\rangle _{s}=\left|
0\right\rangle _{r}\left| 0\right\rangle _{s}\text{ and }A_{12}\left|
1\right\rangle _{r}\left| 1\right\rangle _{s}=\left| 1\right\rangle
_{r}\left| 1\right\rangle _{s}, 
\]
without forgetting that $\left| 0\right\rangle _{r}\left| 0\right\rangle
_{s}=\left| c\right\rangle $ and $\left| 1\right\rangle _{r}\left|
1\right\rangle _{s}=\left| d\right\rangle $ are {\em excited states}.

The link can be implemented by suitably operating on the ground state (6).
We assume this to be initially given by eq. (4), repeated here:

\[
\left| \Psi \left( 0\right) \right\rangle =\cos \vartheta \left|
0\right\rangle _{r}\left| 1\right\rangle _{s}+\sin \vartheta \left|
1\right\rangle _{r}\left| 0\right\rangle _{s}. 
\]
The transformation of qubit $r$ under (the equivalent of) a continuous $%
A_{rs}$ projection is described by:

\noindent {\em for all }$t$:

\begin{enumerate}
\item[i)]  $A_{12}\left| \Psi \left( t\right) \right\rangle =\left| \Psi
\left( t\right) \right\rangle ,$

\item[ii)]  $Max$ $\left| \left\langle \Psi \left( t\right) \right| \left.
\Psi \left( t-dt\right) \right\rangle \right| ,$

\item[iii)]  $diag$ $\rho _{r}\left( t\right) =diag\left\{ Tr_{s}\left[
\left( \left| \Psi \left( t\right) \right\rangle \left\langle \Psi \left(
t\right) \right| \right) \right] \right\} =$
\end{enumerate}

$\cos ^{2}\left[ \vartheta +\varphi \left( t\right) \right] \left|
0\right\rangle _{r}\left\langle 0\right| _{r}+\sin ^{2}\left[ \vartheta
+\varphi \left( t\right) \right] \left| 1\right\rangle _{r}\left\langle
1\right| _{r},$

\begin{enumerate}
\item[iv)]  $\left\langle \xi _{rs}\left( t\right) \right\rangle
=\left\langle \Psi \left( t\right) \right| H_{rs}\left| \Psi \left( t\right)
\right\rangle =0,$

\noindent where $\left| \Psi \left( t\right) \right\rangle $ is a free
normalized vector of ${\cal H}_{12}$.
\end{enumerate}

The solution of the above equations is the desired link evolution (5),
repeated here:

\[
\left| \Psi \left( t\right) \right\rangle =\cos \left[ \vartheta +\varphi
\left( t\right) \right] \left| 0\right\rangle _{r}\left| 1\right\rangle
_{s}+\sin \left[ \vartheta +\varphi \left( t\right) \right] \left|
1\right\rangle _{r}\left| 0\right\rangle _{s}. 
\]

\noindent \noindent Conditions (i), (ii) and (iii) mean that the link state
undergoes a transformation [driven by (iii)] under continuous state vector
reduction on the {\em antisymmetric} subspace ${\cal H}_{12}^{\left(
a\right) }$. If $\vartheta \neq 0,\frac{\pi }{2}$, namely if the preparation
(4) comprises {\em all} the basis vectors of ${\cal H}_{rs}^{\left( c\right)
}$, condition (ii) alone keeps the link evolution inside ${\cal H}%
_{rs}^{\left( c\right) }$ (Section III). The link state remains ground and
consequently the link expected energy $\left\langle \xi _{rs}\left( t\right)
\right\rangle $ is always zero. By excluding $\vartheta =0,\frac{\pi }{2}$,
condition (iv) is a {\em consequence} of the former conditions.

Mathematically, conditions (i) and (iv) give the constraint $A_{rs}\left|
\Psi \left( t\right) \right\rangle =\left| \Psi \left( t\right)
\right\rangle $. In conclusion the above conditions (i) through (iv) \
(which imply interpreting fermionic antisymmetry $A_{12}$ as {\em continuous}
{\em projection} on the antisymmetric subspace) are equivalent to condition
(i) through (iii) of Section III. This gives in fact the evolution (5).

Let us see how we could operate (in principle) on the state of qubit $r$, in
order to implement the driving condition (iii). Since conditions (i) through
(iv) generate (under the counterfactual interpretation of particle
statistics) the projector $A_{rs},$ operating on qubits $r$ and $s$, the
state of qubit $s$ is redundant with respect to the state of qubit $r$. In
particular, operating only on qubit $r$ does not mean applying the identity
to qubit $s$, it means disregarding it (we are now inside that ``peculiar
way of applying quantum mechanics'' highlighted in\ Section II).

In order to operate on qubit $r$, we introduce the one-qubit Hamiltonian $%
H_{r}=E_{r}a_{0r}^{\dagger }a_{0r}$ of ground state $\left| 1\right\rangle
_{r}$ and excited state $\left| 0\right\rangle _{r}$. $H_{r}$ and $H_{rs}$
commute and the total expected energy of the system is $\left\langle \Psi
\left( t\right) \left| \left( H_{rs}+H_{r}\right) \right| \Psi \left(
t\right) \right\rangle $. This is conveniently split into a link internal
energy $\left\langle \xi _{rs}\left( t\right) \right\rangle =\left\langle
\Psi \left( t\right) \left| H_{rs}\right| \Psi \left( t\right) \right\rangle 
$ and a qubit $r$ internal energy $\left\langle \xi _{r}\left( t\right)
\right\rangle =\left\langle \Psi \left( t\right) \left| H_{r}\right| \Psi
\left( t\right) \right\rangle =E_{r}\cos ^{2}\left[ \vartheta +\varphi
\left( t\right) \right] $. One can see that the latter is in one-to-one
correspondence with $diag$ $\rho _{r}\left( t\right) $: the driving
condition (iii) can thus be implemented by changing the internal energy of
qubit $r$, provided that $\left\langle \xi _{rs}\left( t\right)
\right\rangle =0$ (see further below).

Qubit $r$ is now put in interaction with a heat-bath. This could also excite
the link, yielding $\left\langle \xi _{rs}\left( t\right) \right\rangle
=\left\langle \Psi \left( t\right) \left| H_{rs}\right| \Psi \left( t\right)
\right\rangle >0$. In order to take this possibility into account, it is
convenient to split $\left| \Psi \left( t\right) \right\rangle $ into two
orthogonal components: 
\[
\left| \Psi \left( t\right) \right\rangle =\alpha \left( t\right) \left|
\Psi ^{\prime }\left( t\right) \right\rangle +\beta \left( t\right) \left|
\Psi ^{\prime \prime }\left( t\right) \right\rangle , 
\]
with $\left| \alpha \left( t\right) \right| ^{2}+\left| \beta \left(
t\right) \right| ^{2}=1$, such that $\left\langle \Psi ^{\prime }\left(
t\right) \left| H_{rs}\right| \Psi ^{\prime }\left( t\right) \right\rangle
=0,$ $\left\langle \Psi ^{\prime \prime }\left( t\right) \left|
H_{rs}\right| \Psi ^{\prime \prime }\left( t\right) \right\rangle
=\left\langle \xi _{rs}\left( t\right) \right\rangle >0$, $\left| \Psi
^{\prime }\left( t\right) \right\rangle $ belongs to ${\cal H}_{rs}^{\left(
c\right) }$ and $\left| \Psi ^{\prime \prime }\left( t\right) \right\rangle $
belongs to the orthogonal subspace.

The former component is hosted, so to speak, in a ``good'' Everett universe
where $\left\langle \xi _{rs}\left( t\right) \right\rangle =0$, $A_{rs}$ is
turned on and the link works as required: the transformation performed on
qubit $r$ drags a corresponding transformation of qubit $s$. From the
standpoint of the link internal energy, this occurs in a reversible way with
no free energy dissipation (nor increase).

The latter component is hosted in a ``bad'' Everett universe where the link
state is a linear combination of the excited states $\left| 0\right\rangle
_{r}\left| 0\right\rangle _{s}$ and $\left| 1\right\rangle _{r}\left|
1\right\rangle _{s}$ and the link (or NOT\ gate) logical constraint is not
satisfied. \ 

$\beta \left( 0\right) $ was zero in the preparation (4). Let $T$ be the
heat-bath temperature. We assume that $k_{B}T<<E\leqslant E_{a}$, $E_{b}$,
etc.. Thus $\beta \left( t\right) $ will remain close to zero after the
heat-bath has been turned on. Therefore, the probability $\left| \alpha
\left( t\right) \right| ^{2}$ that measurement finds the link in the
``good'' Everett universe should remain very high.

We further assume that $E_{r}<<E$ is of the order of $k_{B}T$. $\left\langle
\xi _{r}\left( t\right) \right\rangle $ is driven to relax close to zero by
reducing $T$. At a time $\tau $ such that $\left\langle \xi _{r}\left( \tau
\right) \right\rangle =\cos ^{2}\left[ \vartheta +\varphi \left( \tau
\right) \right] \simeq 0$, and in the ``good'' Everett universe, the link
state has the form $\left| \Psi ^{^{\prime }}\left( \tau \right)
\right\rangle \simeq \left| 1\right\rangle _{r}\left| 0\right\rangle _{s}$.
The time required to reach $\left\langle \xi _{r}\left( \tau \right)
\right\rangle \simeq 0$ is uniquely determined by the relaxation process of
qubit $r$ -- independently of the rest of the link.

This way of driving the link evolution, whose plausibility is based on
heuristics for the time being, will be used to drive the evolution of the
entire network (Section VI).

\section{The generic gate}

To reduce notation, we shall work with a gate of three (coexisting) qubits,
the minimum required to have a conditional logic. The result will be clearly
generalizable to any number of qubits. We will consider the logically
irreversible XOR gate, of inputs $t,u$, output $v$, and thruth table:

\begin{center}
\begin{tabular}{|c|c|c|}
\hline
\multicolumn{2}{|c}{\small I} & \multicolumn{1}{|c|}{\small O} \\ \hline
\multicolumn{1}{|c|}{$t$} & $u$ & $v$ \\ \hline
$0$ & $0$ & $0$ \\ \hline
$0$ & $1$ & $1$ \\ \hline
$1$ & $0$ & $1$ \\ \hline
$1$ & $1$ & $0$ \\ \hline
\end{tabular}
\end{center}

Since time is orthogonal to the network lay-out, the network evolution can
be reversible even if gates are {\em logically} irreversible.

The gate mathematical model will be introduced first. Let

\[
{\cal H}_{tuv}=span\left\{ \left| 0\right\rangle _{t}\left| 0\right\rangle
_{u}\left| 0\right\rangle _{v},\left| 0\right\rangle _{t}\left|
0\right\rangle _{u}\left| 1\right\rangle _{v},...,\left| 1\right\rangle
_{t}\left| 1\right\rangle _{u}\left| 1\right\rangle _{v}\right\} 
\]

\noindent be the eight-dimensional Hilbert space of the gate three qubits.
The constrained subspace (whose basis vectors map the rows of the gate truth
table) is four-dimensional:

\[
{\cal H}_{tuv}^{\left( c\right) }=span\left\{ \left| 0\right\rangle
_{t}\left| 0\right\rangle _{u}\left| 0\right\rangle _{v},\left|
0\right\rangle _{t}\left| 1\right\rangle _{u}\left| 1\right\rangle
_{v},\left| 1\right\rangle _{t}\left| 0\right\rangle _{u}\left|
1\right\rangle _{v},\left| 1\right\rangle _{t}\left| 1\right\rangle
_{u}\left| 0\right\rangle _{v}\right\} ; 
\]

\noindent The projector $A_{tuv}$ from ${\cal H}_{tuv}$ on ${\cal H}%
_{tuv}^{\left( c\right) }$ is defined by the (eight) equations:

\medskip

$A_{tuv}\left| \chi _{\alpha }\right\rangle _{t}\left| \chi _{\beta
}\right\rangle _{u}\left| \chi _{\gamma }\right\rangle _{v}=\left| \chi
_{\alpha }\right\rangle _{t}\left| \chi _{\beta }\right\rangle _{u}\left|
\chi _{\gamma }\right\rangle _{v}$ when $\stackrel{t}{\chi _{\alpha }}%
\stackrel{u}{\chi _{\beta }}\stackrel{v}{\chi _{\gamma }}$ is a row of the
gate truth table,

\medskip

$A_{tuv}\left| \chi _{\alpha }\right\rangle _{t}\left| \chi _{\beta
}\right\rangle _{u}\left| \chi _{\gamma }\right\rangle _{v}=0$ when $%
\stackrel{t}{\chi _{\alpha }}\stackrel{u}{\chi _{\beta }}\stackrel{v}{\chi
_{\gamma }}$ is not a row of the gate truth table (like e.g. $\stackrel{t}{0}%
\stackrel{u}{1}\stackrel{v}{0}$).

Let us denote by $\left| \Psi \left( t\right) \right\rangle $ the evolution
of the gate state. $\left| \Psi \left( t\right) \right\rangle $ is
continuously projected on ${\cal H}_{tuv}^{\left( c\right) }$. This is
represented by:

\medskip

\noindent for all $t:$

\begin{enumerate}
\item[i)]  $A_{tuv}\left| \Psi \left( t\right) \right\rangle =\left| \Psi
\left( t\right) \right\rangle ,$

\item[ii)]  $Max$ $\left| \left\langle \Psi \left( t\right) \right. \left|
\Psi \left( t-dt\right) \right\rangle \right| ,$

\noindent where $\left| \Psi \left( t\right) \right\rangle $ is an otherwise
free normalized vector of $H_{tuv}$; in fact conditions (i) and (ii) do not
yet determine $\left| \Psi \left( t\right) \right\rangle $: the ``missing''
driving condition (iii) will come out from the behaviour of the rest of the
network (Section VI).
\end{enumerate}

\noindent The extension to a gate with any number of qubits is obvious.

Now we should build the gate model Hamiltonian $H_{tuv}$. We need three
lattice sites $t,u,v$ and three identical spin $\frac{1}{2}$ particles $%
1,2,3 $. \ The unwanted states should all be excited eigenstates of $H_{tuv}$%
. This has therefore the form: 
\[
H_{tuv}=...-E_{\nu }a_{0t}^{\dagger }a_{1t}^{\dagger }a_{0v}^{\dagger
}a_{0t}a_{1t}a_{0v}\text{ }...\text{ }-E_{\gamma }a_{0t}^{\dagger
}a_{1u}^{\dagger }a_{0v}^{\dagger }a_{0t}a_{1u}a_{0v}\text{ }... 
\]

\noindent The first term corresponds to a state where each site is not
occupied by exactly one particle. The second term corresponds to a state
where each site is occupied by exactly one particle but is not a basis
vector of ${\cal H}_{tuv}^{\left( c\right) }$, etc..

\bigskip Conditions (i), (ii) and (iv) of Section IV become now:

\bigskip for all $t$:

\begin{enumerate}
\item[i)]  $A_{123}\left| \Psi \left( t\right) \right\rangle =\left| \Psi
\left( t\right) \right\rangle $, where $A_{123}=\frac{1}{6}\left(
1-P_{12}-P_{13}-P_{23}+P_{13}P_{12}+P_{23}P_{12}\right) $ is the
antisymmetrization projector,

\item[ii)]  $Max\left| \left\langle \Psi \left( t\right) \right. \left| \Psi
\left( t-dt\right) \right\rangle \right| ,$

\item[iv)]  $\left\langle \xi _{tuv}\left( t\right) \right\rangle
=\left\langle \Psi \left( t\right) \right| H_{tuv}\left| \Psi \left(
t\right) \right\rangle =0,$

\noindent where $\left| \Psi \left( t\right) \right\rangle $ is an otherwise
free normalized vector of ${\cal H}_{123}$ (similar to ${\cal H}_{12}$ of
Section IV).
\end{enumerate}

\noindent\ \ The gate initial state should be a linear combination of all $%
{\cal H}_{tuv}^{\left( c\right) }$ basis vectors. Consequently, because of
(ii), the gate state will remain in ${\cal H}_{tuv}^{\left( c\right) }$:
condition (iv) is thus a consequence of (i) and (ii). (i) and (iv) are
equivalent to $A_{tuv}\left| \Psi \left( t\right) \right\rangle =\left| \Psi
\left( t\right) \right\rangle $. $\ A_{tuv}$ is thus ``turned on'' when $%
\left\langle \xi _{tuv}\left( t\right) \right\rangle =0$.

\section{The network}

We must put together the stand-alone network elements of the previous\
Sections. The network lattice is the union of the lattices of the individual
N.E. Of course not all sites are disjoint: for example a gate and a
connected link share one lattice site. All N.E. \ Hamiltonians $H_{i}$
(where $i$ is a multiple index running over $de$, $abcd$, etc. -- fig. 1)
are diagonal in\ ${\cal H}_{N}$ basis thus pairwise commuting: $\forall
_{i,j}:\left[ H_{i},H_{j}\right] =0.$

The overall network Hamiltonian is: $H_{N}=\sum_{i}H_{i},$ and the network
expected energy $\left\langle \xi _{N}\left( t\right) \right\rangle $ is the
sum of the expected energies of all N.E.: $\left\langle \xi _{N}\left(
t\right) \right\rangle =\sum_{i}\left\langle \Psi \left( t\right) \right|
H_{i}\left| \Psi \left( t\right) \right\rangle ,$ where $\left| \Psi \left(
t\right) \right\rangle $ is the network state (i.e. a linear combination of
the network basis vectors -- each a tensor product of all qubit eigenstates).

Therefore, if $\left| \Psi \left( t\right) \right\rangle $ is a nework
ground state (this is highly degenerate, see Section I), then $\left\langle
\xi _{N}\left( t\right) \right\rangle =0$ and consequently $\left\langle
\Psi \left( t\right) \right| H_{i}\left| \Psi \left( t\right) \right\rangle
=0$ for all $i$. Fermion statistics and these latter conditions work as
conditions (i) and (iv) of Section V: all $A_{i}$ projectors are ``turned
on'' when $\left\langle \xi _{N}\left( t\right) \right\rangle =0$.

Of course $\forall _{i,j}:\left[ A_{i},A_{j}\right] =0$. $A_{N}=\Pi
_{i}A_{i} $ projects ${\cal H}_{N}$ on the constrained subspace ${\cal H}%
_{N}^{\left( c\right) }$ spanned by the network basis vectors which satisfy 
{\em all} N.E. (${\cal H}_{N}^{\left( c\right) }$ is the intersection of all 
${\cal H}_{i}^{\left( c\right) }$)\footnote{%
We should note that the permutation of two identical particles belonging to
any two different N.E. cannot set any further particle statistics
constraint, since the two N.E. Hamiltonians commute.}.

To sum up, any transformation of the network ground state $\left| \Psi
\left( t\right) \right\rangle $ satisfies the following conditions:

for all $t:$

\begin{enumerate}
\item[i)]  $A_{N}\left| \Psi \left( t\right) \right\rangle =\left| \Psi
\left( t\right) \right\rangle ,$

\item[ii)]  $Max$ $\left| \left\langle \Psi \left( t\right) \right| \left.
\Psi \left( t-dt\right) \right\rangle \right| ,$

\item[iv)]  $\left\langle \xi _{N}\left( t\right) \right\rangle =0,$

\noindent where $\left| \Psi \left( t\right) \right\rangle $ is an otherwise
free vector of ${\cal H}_{N}$.
\end{enumerate}

\noindent Of course, any tensor product of the network qubit eigenstates
satisfying condition (i) satisfies all the N.E., not necessarily the input
and output constraints.

The satisfiability problem requires checking whether such constraints can
also be satisfied. Under the assumption that the network is satisfiable
together with the input-output constraints, we proceed as follows.

\bigskip

1) We assume, with no restriction, that only one output qubit, say $z$, is
constrained: $\rho _{z}=\left| 1\right\rangle _{z}\left\langle 1\right| _{z}$%
. This output constraint is temporarily removed. Whereas we introduce
one-qubit Hamiltonians in order to detain the constrained inputs in the
required values. Let $h$ be an input qubit constrained by $\rho _{h}=\left|
0\right\rangle _{h}\left\langle 0\right| _{h}$ ($\rho _{h}=\left|
1\right\rangle _{h}\left\langle 1\right| _{h}$). This constraint is
satisfied in the ground state of the one-qubit Hamiltonian $%
H_{h}=E_{h}a_{1h}^{\dagger }a_{1h}$ ($H_{h}=E_{h}a_{0h}^{\dagger }a_{0h}$),
with $E_{h}$ discretely above zero (of the same order of $E$). We should
note that these Hamiltonians commute with all the other N.E. Hamiltonians.

\bigskip

2) We wish to prepare the network in a state $\left| \Psi \left( 0\right)
\right\rangle $ which is a linear combination of {\em all} ${\cal H}_{N}$
basis vectors satisfying the input constraints, all the N.E., but not
necessarily the output constraint (not implemented until now). This means
satisfying the equations:

\begin{itemize}
\item  $\forall $ $h:\left\langle \Psi \left( 0\right) \right| H_{h}\left|
\Psi \left( 0\right) \right\rangle $, where $h$ ranges over the labels of
the input qubit and N.E. Hamiltonians;

\item  $\forall $ $k:\rho _{k}\left( 0\right) =a_{k}\left| 0\right\rangle
_{k}\left\langle 0\right| _{k}+b_{k}\left| 1\right\rangle _{k}\left\langle
1\right| _{k},$ with $a_{k}$, $b_{k}\neq 0$, where $k$ ranges over the
labels of the unconstrained input qubits; we should note that the coherence
elements of any qubit density matrix are zero if, without any restriction,
any two qubits are connected by a link (although this is not the case of
fig.1).
\end{itemize}

The above conditions yield:

\begin{equation}
\left| \Psi \left( 0\right) \right\rangle =\cos \vartheta \sum_{i}\alpha
_{i}\left| prep_{i}\right\rangle \left| 0\right\rangle _{z}+\sin \vartheta
\sum_{j}\beta _{j}\left| prep_{j}\right\rangle \left| 1\right\rangle _{z}.
\end{equation}
with $\sum_{i}\left| \alpha _{i}\right| ^{2}=\sum_{j}\left| \beta
_{j}\right| ^{2}=1;$ $\left| prep_{i}\right\rangle $, $\left|
prep_{j}\right\rangle $ denote tensor products of all qubit eigenstates but
qubit $z$.

The ground state (7) can be reached through quantum annealing, namely
quantum ground state computation$^{\left[ 11,19\right] }$. Since there is no
output constraint, this amounts to solving a problem polynomial in network
size. Therefore, {\em for short}, without entering into any detail, we
assume that the time required to reach any desired probability of finding
the network in the ground state (7) is polynomial in network size.

In the right side of eq. (7), each term $i$ satisfies the input constraints,
all N.E., {\em not} the output constraint, whereas each term $j$ satisfies
also the output constraint and is therefore a network solution.

$\sin ^{2}\vartheta $ depends on the network and must decrease exponentially
with network size, otherwise the problem would not be hard (solutions are
``exponentially \ rare''). Of course:

\[
diag\rho _{z}\left( 0\right) =\cos ^{2}\vartheta \left| 0\right\rangle
_{z}\left\langle 0\right| _{z}+\sin ^{2}\vartheta \left| 1\right\rangle
_{z}\left\langle 1\right| _{z}, 
\]

with $\sin ^{2}\vartheta $ ``extremely small''.

\bigskip

3) We operate now on qubit $z$ as we did in Section IV, so that at some time 
$\tau $: 
\[
diag\rho _{z}\left( \tau \right) \simeq \left| 1\right\rangle
_{z}\left\langle 1\right| _{z}.
\]
This is done by introducing the one-qubit Hamiltonian $H_{z}=E_{z}a_{z0}^{%
\dagger }a_{z0}$, with $E_{z}<<E$ and ground state $\left| 1\right\rangle
_{z}$, and by putting the qubit in interaction with a heat-bath whose
temperature is lowered down close to zero. The internal energy of qubit $z$
is thus $\left\langle \xi _{z}\left( t\right) \right\rangle =$ $\left\langle
\Psi \left( t\right) \right| H_{z}\left| \Psi \left( t\right) \right\rangle $%
. \ At some time $t=\tau $ (``after'' relaxation), we have: $\left\langle
\xi _{z}\left( \tau \right) \right\rangle \simeq 0$.

\hspace{0in}\noindent As in Section IV, the network evolution can be split
into two orthogonal components: $\left| \Psi \left( t\right) \right\rangle
=\alpha \left( t\right) \left| \Psi ^{^{\prime }}\left( t\right)
\right\rangle +\beta \left( t\right) \left| \Psi ^{^{\prime \prime }}\left(
t\right) \right\rangle $;

\begin{itemize}
\item  $\left| \Psi ^{^{\prime }}\left( t\right) \right\rangle $ is ground
(with reference to the network and the constrained input qubits internal
energy, see Section IV), namely $\sum_{h}\left\langle \Psi ^{^{\prime
}}\left( t\right) \right| H_{h}\left| \Psi ^{^{\prime }}\left( t\right)
\right\rangle =0$, with $h\neq z$ ranging over all the N.E. and the
constrained input Hamiltonians; this state is hosted in the ``good''\
Everett universe of amplitude $\alpha \left( t\right) $;

\item  $\left| \Psi ^{^{^{\prime \prime }}}\left( t\right) \right\rangle $
is excited and belongs to the subspace orthogonal to ${\cal H}_{N}^{\left(
c\right) }$; it is hosted in the ``bad'' Everett universe of amplitude $%
\beta \left( t\right) $.
\end{itemize}

\hspace{0in}We assume to be in the ``good'' Everett universe (until
different notice). Qubit $z$ relaxation, under the continuous projection (i)
and (ii) ${\em drives}$ the transformation of the overall network state: 
\[
\left| \Psi \left( t\right) \right\rangle =\cos \left[ \vartheta +\varphi
\left( t\right) \right] \sum_{i}\alpha _{i}\left| prep_{i}\right\rangle
\left| 0\right\rangle _{z}+\sin \left[ \vartheta +\varphi \left( t\right) %
\right] \sum_{j}\beta _{j}\left| prep_{j}\right\rangle \left| 1\right\rangle
_{z}; 
\]

\noindent The driving condition is $E_{z}\cos ^{2}\left[ \vartheta +\varphi
\left( t\right) \right] =\left\langle \xi _{z}\left( t\right) \right\rangle $%
; we should keep in mind that, in current assumptions, qubit $r$ relaxation
is independent of the rest of the network. Condition (ii) keeps all $\alpha
_{i}$ and $\beta _{j}$ unaltered throughout the evolution, as readily
checked. At time $t=\tau $:

\[
\left| \Psi ^{^{\prime }}\left( \tau \right) \right\rangle \simeq
\sum_{j}\beta _{j}\left| prep_{j}\right\rangle \left| 1\right\rangle _{z}. 
\]

\noindent \hspace{0in}By measuring the network at time $\tau $, and by
repeating the overall process for a sufficient number of times, a
straightforward application of probability theory shows that the number of
repetitions required to check whether the network is satisfiable (with any
desired probability that this check is correct) grows polynomially with
network size. Furthermore, $\tau $ is independent of network size. This
would mean NP-complete = P.

\hspace{0in}However\noindent \hspace{0in}, this result can be completely
vanified if the probability of finding the network in the ``good'' universe
decreases exponentially with network size. To find out how this probability
decreases, we should compute how the qubit $r$-heat-bath interaction affects 
$\alpha \left( t\right) $ and $\beta \left( t\right) $. Doing this would
require a level of formalization beyond the reach of the current study.

Anyway, one ingredient of the ``quantum computation speed-up'' consists of
substituting interaction, which requires time, with interference, which
requires no time. Particle statistics, in the current interpretation, works
as interference and implements most of the network logical constraints.
This, in a still generic way, should justify the hope of achieving a
computation speed-up by using particle statistics.

\section{Discussion}

We have highlighted an alternative way of approaching quantum computation.
With respect to time-sequential computation (Section I), it presents some
significant differences:

\begin{itemize}
\item  logical constraints are simultaneous in time rather than being mapped
on the time evolution of the computation process;

\item  there is thus an analogy between such constraints and the
simultaneous constraints (also logical in character) established by particle
statistics;

\item  in order to make this analogy explicit, we must introduce a blunt
interpretation of particle statistics. A particle statistics symmetry would
no more be a passive constant of motion which does nothing to a unitary
evolution, but an active watchdog effect shaping that evolution;

\item  the NP-complete SAT problem is in a way native in this approach.
\end{itemize}

The counterfactual interpretation of particle statistics might be
interesting in itself. Checking it might not be out of reach. The link
behaviour (Section IV) is similar to exchange interaction. One should devise
an experimental situation, showing that an operation on one part drags the
state of the other part according to the formulation propounded.

We should finally note that this formulation of particle statistics can be
represented in a two-way (advanced and retarded) propagation model. This has
been done in ref. [12] and can be outlined as follows. We have been dealing
with state vector reduction on a {\em predetermined} subspace, namely always
on the constrained subspace ${\cal H}^{\left( c\right) },$ never on the
orthogonal subspace violating the constraint. Although elusively (since we
are in counterfactual reasoning), this means an evolution affected by a
condition coming from the future.

Interestingly, a two-way propagation model of partial state vector reduction
can also justify the speed-up of time-sequential computation. See ref. [12].

This research has been partly developed during the Elsag Bailey-ISI Workshop
on Quantum Computation (Turin, 1997). Thanks are due to A. Ekert, D.
Finkelstein, S. Lloyd and V. Vedral for useful suggestions.

\end{document}